\documentclass[12pt]{article}
\usepackage{latexsym}

\begin{document}

\title{Comparing Different Improvement Programs for the $N$-Vector Model}
\author{
  \\
  {\small Sergio Caracciolo}\thanks{
  Address after November 1:
  Scuola Normale Superiore, I-56100 Pisa, Italy.}               \\[-0.2cm]
  {\small\it Dipartimento di Fisica and INFN -- Sezione di Lecce}  \\[-0.2cm]
  {\small\it Universit\`a degli Studi di Lecce}        \\[-0.2cm]
  {\small\it I-73100 Lecce, ITALIA}          \\[-0.2cm]
  {\small e-mail: {\tt caraccio@sns.it}}     \\[-0.2cm]
  \\[-0.1cm]  \and
  {\small Andrea Pelissetto}              \\[-0.2cm]
  {\small\it Dipartimento di Fisica and INFN -- Sezione di Pisa}    \\[-0.2cm]
  {\small\it Universit\`a degli Studi di Pisa}        \\[-0.2cm]
  {\small\it I-56100 Pisa , ITALIA}          \\[-0.2cm]
  {\small e-mail: {\tt pelisset@ibmth1.difi.unipi.it}}   \\[-0.2cm]
  {\protect\makebox[5in]{\quad}}  
  \\
}
\vspace{0.5cm}

\maketitle
\thispagestyle{empty}   

\vspace{0.2cm}

\begin{abstract}
We discuss the connection between various types of improved actions
in the context of the two-dimensional $\sigma$-model.  We also discuss 
spectrum-improved actions showing that these actions do not have 
any improved behaviour. 
An $O(a^2)$ on-shell improved action with all couplings 
defined on a plaquette and satisfying reflection positivity 
is also explicitly constructed. 
\end{abstract}

\clearpage

\newcommand{\be}{\begin{equation}}
\newcommand{\ee}{\end{equation}}
\newcommand{\<}{\langle}
\renewcommand{\>}{\rangle}

\def\spose#1{\hbox to 0pt{#1\hss}}
\def\ltapprox{\mathrel{\spose{\lower 3pt\hbox{$\mathchar"218$}}
 \raise 2.0pt\hbox{$\mathchar"13C$}}}
\def\gtapprox{\mathrel{\spose{\lower 3pt\hbox{$\mathchar"218$}}
 \raise 2.0pt\hbox{$\mathchar"13E$}}}

\def\bsigma{\mbox{\protect\boldmath $\sigma$}}
\def\bpi{\mbox{\protect\boldmath $\pi$}}
\def\btau{\mbox{\protect\boldmath $\tau$}}
\def\hatp{\hat p}
\def\hatl{\hat l}

\def\msbar{ {\overline{\hbox{\scriptsize MS}}} }
\def\normalmsbar{ {\overline{\hbox{\normalsize MS}}} }

\newcommand{\R}{\hbox{{\rm I}\kern-.2em\hbox{\rm R}}}

\newcommand{\reff}[1]{(\ref{#1})}
\def\smfrac#1#2{{\textstyle\frac{#1}{#2}}}


In recent years there has been much work in improving lattice actions.
The idea behind all these attempts is to modify the lattice 
action with the addition of irrelevant operators in order to reduce lattice
artifacts: in this way one hopes  to have scaling (and finite-size
scaling) at smaller correlation lengths.

There have been many different approaches to the problem of 
improving lattice actions. In this letter  we will address the problem in the 
context of two-dimensional $N$-vector models, trying to point out differences 
and similarities of the various approaches.

The first systematic study of improvement of lattice actions is due to 
Symanzik \cite{Symanzik}. 
The idea is very simple. Consider on a square lattice the standard action 
\be
S^{std} = N \sum_{x\mu} \bsigma_x\cdot\bsigma_{x+\mu}
\ee
where the fields $\bsigma_x$ are $N$-component spins; 
the partition function is
\be
Z = \int \prod_x d\bsigma_x \delta(\bsigma_x^2 - 1)\, e^{\beta S} \;\; .
\ee
Then consider the 
one-particle irreducible Green's functions 
$\Gamma^{(n)}(p_1,\ldots,p_n)$. It is easy to see that at tree level
\be
\Gamma^{(n)}(p_1,\ldots,p_n) = 
\Gamma^{(n)}_{cont}(p_1,\ldots,p_n) + O(a^2)
\label{eq1}
\ee
where $\Gamma^{(n)}_{cont}(p_1,\ldots,p_n)$ is the 
{\em continuum} $n$-point function.

The strategy proposed by Symanzik consists in modifying the action so as
to cancel the terms of $O(a^2)$ in \reff{eq1}. The simplest action 
which satisfies this condition at tree-level is the improved action
\be
S^{Sym} = 
     N \sum_{x\mu} \left( {4\over3} \bsigma_x \cdot \bsigma_{x+\mu} \, -\, 
         {1\over12} \bsigma_x \cdot \bsigma_{x+2\mu} \right) \;\; .
\label{SSymanzik}
\ee
In the Symanzik approach one can proceed further in two different directions:
first of all one can improve the action to order $O(a^4)$, $O(a^6)$ and 
so on. This does not seem to be particularly interesting: indeed,
even if the action is tree-level improved to order $O(a^{2k})$, $k>1$,
then corrections of order $O(a^2)$ will again appear at one- and higher-loop
order.
The second important characteristic of the Symanzik approach is that it can 
be systematically extended to higher loops: in other words 
one can remove terms of order $O(g^2 a^2)$, $O(g^4 a^2)$ and so 
on within perturbation theory.

A second approach to the improvement of lattice actions (the so-called 
on-shell improvement) has been put forward by L\"uscher and Weisz 
\cite{LW}.
The idea here is to improve only spectral quantities like the masses 
of stable particles. In the $O(N)$ $\sigma$-model,
in a strip $L\times\infty$, 
one can consider the mass gap $\mu(\beta,L)$ defined by
\be 
\mu(\beta,L) = - \lim_{y\to +\infty} {1\over y} 
     \log \left[ \sum_{x=1}^L\< \bsigma_0\cdot \bsigma_{(x,y)}\>\right]
\;\; .
\ee
Then one considers the asymptotic expansion of $\mu(\beta,L)$ for 
$\beta\to\infty$ at $L$ fixed which will have generically the form
\be
\mu(\beta,L) L\, =\, \sum_{n=1}^\infty {\mu_n(L)\over \beta^n} \;\; .
\label{mupert}
\ee
Each coefficient $\mu_n(L)$ has an expansion in powers of $1/L^2$ 
(with additional logarithms of $L$). The $O(a^2)$ improved action is then 
chosen by requiring, order by order in perturbation theory,
that $\mu_n(L)$ does not have $1/L^2$ corrections.
It should be noticed that $O(a^2)$ improvement of the one-loop term
$\mu_2(L)$ gives the $O(a^2)$ tree-level improved action \cite{LW}.

Let us now discuss the two different approaches in the soluble case of 
$N=\infty$. 
Let us consider the generalized action
\be
S =\, N \sum_{x,y} J(x-y) \bsigma_x \cdot \bsigma_y \;\; .
\label{eq2}
\ee
We will assume the interaction to be local and parity invariant:
if $\widehat{J}(p)$ is
the Fourier transform of $J(x)$, we will require that
$\widehat{J}(p)$ is a continuous function of $p$, even under $p\to -p$.
We will also
require invariance under rotations of $\pi/2$, that is we will
assume that $\widehat{J}(p)$ is symmetric under exchange of ${p}_1$ and
${p}_2$. Redefining $\beta$ we can normalize the
couplings so that
\be
\widehat{J}(q) = \widehat{J}(0) - {q^2 \over 2} + O(q^4)
\ee
for $q\to 0$. We also introduce the function
\be
    w(q) = - 2 (\widehat{J}(q) - \widehat{J}(0))
\ee
which behaves as $q^2$ for $q\to 0$. 
Finally we will require the theory to have
the usual (formal) continuum limit: we will assume that
the equation $w(q) = 0$ has only one solution for
$-\pi \le q_i \le \pi$, namely $q=0$.
We will need the small-$q$ behaviour of $w(q)$: we will assume in this limit
the form
\be
w(q) =\, \hat{q}^2 + \alpha_1\, \sum_\mu \hat{q}^4_\mu + 
      \alpha_2 \ (\hat{q}^2)^2 + O(q^6)
\ee
where $\alpha_1$ and $\alpha_2$ are arbitrary constants. Here 
$\hat{q}^2 = \hat{q}^2_1 + \hat{q}^2_2$, $\hat{q}_\mu = 2 \sin (q_\mu/2)$.
Notice that for \reff{SSymanzik} we have $w(q)=q^2 + O(q^6)$ i.e.
$\alpha_1 = {1\over12}$ and $\alpha_2 = 0$.

Let us now study the improvement program \`a la Symanzik. Let us consider 
the two-point function in infinite volume. A trivial calculation gives
\be
\<\bsigma_0\cdot\bsigma_x\> \  \equiv G_V(x) =\ 1 + {1\over \beta} 
    \int {d^2p\over (2\pi)^2} {e^{ipx} - 1\over w(p)} + 
    O(e^{-4 \pi \beta})  \;\; .
\ee
Notice that only the tree-level term appears, all successive ones being
zero. The Fourier transform $\widehat{G}_V(q;\infty)$ for $q\not=0$ is then
given by 
\be
  \widehat{G}_V^{-1}(q;\infty) = \beta w(q) + O(e^{-4 \pi \beta})
\ee
Improvement  \`a la Symanzik requires then $w(q) = q^2 + O(q^6)$ ,
i.e. $\alpha_1 = {1\over 12}$ and $\alpha_2 = 0$. When these two
conditions are satisfied the action is improved to all orders in 
perturbation theory. 

Let us now consider the mass-gap $\mu(\beta,L)$ in a strip and its perturbative 
expansion  \reff{mupert}.
At one loop it is easy to compute \cite{CP_LAT96,CPprep}
\begin{eqnarray}
\mu(\beta,L) L &=& {1\over 2\beta} + 
    {1\over \beta^2}\left[ {1\over 4\pi} \log L + 
   {1\over2} (\overline{F}_{00} + \Lambda_0) \right. 
\nonumber \\
&& \qquad \left . + {\pi\over 144 L^2} (1 - 12 \alpha_1) + O(L^{-4})\right] 
    + O(\beta^{-3})
\end{eqnarray}
where 
\begin{eqnarray}
\overline{F}_{00} &=& {1\over 2\pi} \left(\gamma_E - \log \pi + 
    {1\over2} \log 2 \right) \\
\Lambda_0 &=& \int {d^2p\over (2\pi)^2} \left( {1\over w(p)} - 
      {1\over \hat{p}^2} \right)
\end{eqnarray}
and $\gamma_E \approx 0.577215665$ is the Euler constant.
Perturbative improvement requires the cancellation of the $1/L^2$ term
and thus gives the condition $1-12\alpha_1 = 0$, i.e. 
$\alpha_1 = \smfrac{1}{12}$. 
Now consider the next order. A simple calculation gives
\begin{eqnarray}
\hskip -20pt
\mu_3(L) &=& {1\over 8\pi^2} \log^2 L + {1\over 2\pi} 
   (\overline{F}_{00} + \Lambda_0) \log L + 
   {1\over2} (\overline{F}_{00} + \Lambda_0)^2 
\nonumber \\
&& \hskip -20 pt 
  + {1\over L^2}\left[ {1\over144}(1 -12 \alpha_1) \log L + 
   {\pi\over72} (1 - 12\alpha_1) (\overline{F}_{00} + \Lambda_0)
   + {1\over48} (12 \alpha_1 + 12\alpha_2 - 1)\right] 
\nonumber \\ [-2mm]
&& {}
\end{eqnarray}
Requiring the cancellation of the $1/L^2$ 
term we get $\alpha_1 = \smfrac{1}{12}$
and $\alpha_2 = 0$. Thus only Symanzik tree-level improved 
actions are improved at
this level. We can go further and compute $\mu_4(L)$: for 
$\alpha_1 = \smfrac{1}{12}$
and $\alpha_2 = 0$ the coefficient of $1/L^2$ is 
\be
  {1\over 8} \left( {1\over 96\pi} - \Lambda_1\right)
\ee
where
\be
\Lambda_1 =\, \int {d^2p\over (2\pi)^2} 
     \left[ {1\over w(p)^2} - {1\over (\hat{p}^2)^2} +
     {2\over (\hat{p}^2)^3} \left( \alpha_1 \sum_\mu \hat{p}^4_\mu 
       + \alpha_2 (\hat{p}^2)^2 \right) \right] \;\; .
\label{Lambda1testo}
\ee
Thus improving the three-loop contribution to $\mu(\beta,L)$ gives 
\be
\Lambda_1 = {1\over 96 \pi} \;\; .
\label{conditionLambda1}
\ee
Notice that this condition is global, that is it does not only fix the 
small $q^2$ behaviour of $w(q)$ but it depends on the behaviour of $w(q)$ 
over all the Brillouin zone.

A particular action satisfying \reff{conditionLambda1} is given by 
\begin{eqnarray}
\hskip -7pt
S^{Sym2}  =  N\sum_{x\mu} 
  \left[ \left( {4\over3} + 15 a\right) \bsigma_x\cdot\bsigma_{x+\mu} + \, 
         \left(- {1\over12} - 6 a\right) \bsigma_x\cdot\bsigma_{x+2\mu} + \,
         a \bsigma_x\cdot\bsigma_{x+3\mu} \right] \;\; .
\label{Symanzik2}
\end{eqnarray}
where $a = 0.00836533968(1)$.

A peculiarity of the large-$N$ limit
is the fact that, once \reff{conditionLambda1} is satisfied, 
all subsequent coefficients $\mu_n(L)$ do not have $1/L^2$ corrections:
improvement at three-loops is equivalent to improvement to 
all orders in perturbation theory: Symanzik actions 
satisfying \reff{conditionLambda1} are improved to all perturbative 
orders. Of course this feature will not be true 
for finite values of $N$. 

In conclusion, within Symanzik approach only the two conditions 
$\alpha_1 = {1\over12}$ and $\alpha_2=0$ are required to improve
the action to order $O(a^2)$ and to all orders in perturbation theory.
In the on-shell program instead an additional condition 
(equation \reff{conditionLambda1}) is obtained. Notice that naively 
one would have expected the opposite result since Symanzik approach 
is intended to improve all Green's functions while the on-shell
program considers only spectral quantities. This result is due to the
large $N$ limit~\footnote{We thank Giorgio Parisi who told us that this point was
already known to K.~Symanzik.}.
Indeed, one can see that condition
\reff{conditionLambda1}  appears in the Symanzik approach at order $1/N$. At this
order one should consider, beside the two-point function also the four-point
function. If $\Gamma_{\alpha \beta
\gamma \delta}(p,q,r,s)$ is the one-particle irreducible four-point function, we have
\be
\Gamma_{\alpha \beta \gamma \delta}(p,q,r,s) =
-{1\over N} \left[ \delta_{\alpha \beta} \delta_{\gamma \delta} \Delta(p+q) +
\delta_{\alpha \gamma} \delta_{\beta \delta} \Delta(p+r) +
\delta_{\alpha \delta} \delta_{\gamma \beta} \Delta(p+s) \right] + O(1/N^2)
\ee
where
\be
\Delta^{-1}(p) = {1\over 2} \int {d^2k\over (2 \pi)^2} {1\over \left[ w(k) + m_0^2 
\right] \left[w(p+k) + m_0^2 \right]}
\ee
In the perturbative limit
\be
\Delta^{-1}(p) = {1\over \beta w(p)} \left[ 1 + {1\over 2 \beta} 
\int {d^2 k\over (2 \pi)^2} {w(p) - w(k) - w(p+k) \over w(p+k) w(k)} \right] +
O\left(e^{-4 \pi \beta} \right)
\ee
If $\alpha_1 = {1\over12}$ and  $\alpha_2=0$ we have then~\cite{CPprep}
\be
\Delta^{-1}(p) = {1\over \beta p^2} \left\{ 1 + {1\over \beta} \left[
{1\over 4 \pi} \log {p^2\over 32} - \Lambda_0 + {p^2\over 2} \left( \Lambda_1 -
{1\over 96 \pi}\right)\right] + O(p^4 \log p^2) \right\}
\ee
Thus $O(a^4)$-improvement required the condition \reff{conditionLambda1}.
In the Symanzik approach \reff{conditionLambda1} is thus a one-loop improvement
condition for the four-point function. Notice again the peculiarity of the large $N$
limit: improvement of the one-loop result is enough to improve all orders of
perturbation theory.

We have checked that $\alpha_1 = {1\over12}$,  $\alpha_2=0$ and
\reff{conditionLambda1} are sufficient conditions to improve both the scaling and the
finite-size-scaling  behaviour of the theory. Indeed, consider for instance the ratio 
(in infinite volume)
\be
R(\beta) =\, \left({\xi^{(2)}_V(\beta) \over \xi^{(2)}_T(\beta)}\right)^2
\ee
where
\be
\xi^{(2)}_\#(\beta) = {1\over4} 
      {\sum_x |x|^2 G_\# (x) \over \sum_x  G_\# (x)} 
\ee
and 
\be
G_T(x) = \< \sigma_0^a \sigma_0^b; \sigma_x^a \sigma_x^b \>  \;\; .
\ee
Then an easy calculation gives 
\begin{eqnarray}
R(\beta) &=& 6 - 24 e^{-4\pi(\beta-\Lambda_0)}
    \left[4\pi (12\alpha_1 + 16\alpha_2 - 1)(\beta - \Lambda_0)
  \right. \nonumber \\
  && \qquad \left. - (8 \alpha_1 + 8 \alpha_2 - 1 + 32 \pi\Lambda_1)\right] 
    + O(e^{-8\pi\beta}\beta) \;\; .
\end{eqnarray}
Improvement requires then
\begin{eqnarray}
12 \alpha_1 + 16 \alpha_2 - 1 &=& 0 \;\; ,\\
8\alpha_1 + 8\alpha_2 - 1 + 32 \pi \Lambda_1 &=& 0 \;\; .
\end{eqnarray}
The first term clearly vanishes for the Symanzik action for which 
$\alpha_1 = {1\over12}$ and $\alpha_2 =0$. The second one however 
requires the additional condition \reff{conditionLambda1}.

Let us now consider finite-size scaling. On a strip of width $L$, we have 
computed the corrections to the finite-size-scaling function\footnote{The 
leading term was already computed in \cite{Lu}.} 
for $\mu(\beta,L)$. 
In the limit $L\to\infty$, $\beta\to\infty$ with 
$\mu(\beta,L) L \equiv x$ fixed we find \cite{CP_LAT96,CPprep}
\be
 {\mu(\beta,\infty)\over \mu(\beta,L)} \, =\, 
    f_\mu(x) \left(1 + {\Delta_{\mu,1}(x)\over L^2} \log L + 
    {\Delta_{\mu,2}(x)\over L^2} \right)
\label{FSSformula}
\ee
with corrections of order $\log^2 L/L^4$. 
Cancellation of the terms of 
order $\log L/L^2$ requires $12 \alpha_1 + 16\alpha_2 - 1 = 0$, while 
cancellation of the terms $1/L^2$ requires $\alpha_1 = {1\over12}$,
$\alpha_2 = 0$ and the condition \reff{conditionLambda1}. 
Again \reff{conditionLambda1} is necessary to completely eliminate the
corrections of order $1/L^2$. 

We want to make an additional remark on the on-shell improvement program.
For $N=\infty$ tree-level improvement requires the one-loop contribution
to $\mu(\beta,L)$ to have corrections of order $O(L^{-4})$ and thus 
it requires only $\alpha_1 = {1\over12}$; the second coefficient
$\alpha_2$ is arbitrary to this order. This is a peculiarity 
of the large-$N$ limit. For finite values of $N$, using the results
of \cite{RV,CPprep} , we have
\be
L \mu(\beta,L) = {N-1\over 2N\beta} \left[ 1 + 
    {1\over N \beta} (r_1(L) + (N-2) r_2(L) ) + O(\beta^{-2}) \right]
\label{LmuLpert}
\ee
with 
\begin{eqnarray}
\hskip -10pt
r_1(L) &=& - \int {d^2 k\over (2\pi)^2} {1\over w(k)} 
   \left[ {1\over4} \sum_\mu {\partial^2 w(k)\over \partial k_\mu^2} -
1\right] 
- {\pi\over 12 L^2} (12 \alpha_1 + 8 \alpha_2 -1) + O(L^{-4}) 
\;\; , \nonumber \\ [-2mm]
{} \label{r1L} \\
\hskip -10pt
r_2(L) &=& {1\over 2\pi} \log L 
   + \overline{F}_{00} + \Lambda_0 
   + {\pi\over 72 L^2} (1 - 12\alpha_1) + O(L^{-4})
\;\; .
\label{r2L}
\end{eqnarray}
Clearly the $1/L^2$ corrections cancel only if $\alpha_1 = {1\over12}$
and $\alpha_2 = 0$ except for $N=\infty$.  Let is notice again \cite{LW}
that the one-loop computation fixes the improved action 
at tree-level\footnote{On-shell improvement to order $a^{2k}$, $k>1$,
was discussed in \cite{RV}.
It was shown numerically that if one chooses 
the generalization of \reff{SSymanzik} improved to order 
$O(a^{2k})$, then $\mu_2(L)$ has corrections of order $O(L^{-2k-2})$.
Assuming this result it is trivial to prove rigorously that for
{\em any} action such that $w(q)=q^2 + O(q^{2k+4})$ 
(i.e. for any $O(a^{2k})$ tree-level Symanzik-improved action) 
$\mu_2(L)$ has corrections of order $O(L^{-2k-2})$
(i.e. the action is also $O(a^{2k})$ tree-level on-shell improved in
the sense of L\"uscher and Weisz \cite{LW}). 
Notice that the opposite result is not true: it is {\em not necessary} 
that $w(q)=q^2 + O(q^{2k+4})$ to have $\mu_2(L) = 
\hbox{\rm leading} + O(L^{-2k-2})$. For instance consider 
$w(q)$ such that for $q\to 0$
\be
w(q) = q^2 + \beta_1 \sum_\mu q^8_\mu + 
             \beta_2 (q^2)^2 \sum_\mu q^4_\mu + \beta_3 (q^2)^4  
     + O(q^{10}) \;\; .
\ee
Requiring that $\mu_2(L)$ does not have terms of order $1/L^4$ we get two 
equations (corresponding to $r_1(L)$ and $r_2(L)$) which will not fix 
completely the three parameters $\beta_i$. Thus at order $O(a^6)$ one can have an
on-shell action which is a Symanzik-improved one. }.

We want now to discuss a third type of actions which we will call 
classically spectrum-improved actions. The idea 
which has been put forward by the Bern group
\cite{Niedermayer_LAT96,perfect} consists in changing the action
so that to improve  the dispertion relations.
For instance an action is $O(a^2)$-spectrum improved if the 
equation $w(iE,p)=0$ gives the continuum dispersion relation 
$E^2 - p^2 = 0$ modulo terms of order $O(a^4)$. In practice 
$w(q)$ must have an expansion with $\alpha_1 = {1\over12}$ and $\alpha_2$
arbitrary, i.e.
\be
w(q) = q^2 + \alpha_2 (q^2)^2 + O(q^6) \;\; .
\ee
An example is the action proposed in \cite{Niedermayer_LAT96}:
\be
S^{diag} = 
    N \sum_{x} \left( {2\over3} \sum_\mu \bsigma_x \cdot \bsigma_{x+\mu} 
    \, +\, {1\over6} \sum_{\hat{d}}\bsigma_x \cdot \bsigma_{x+\hat{d}} \right)
\label{Symanzikdiag}
\ee
where $\hat{d}$ are the two diagonal vectors $(1,\pm1)$. For this action
$\alpha_1 = {1\over 12}$ and $\alpha_2 = - {1\over 12}$.

We will now show that these actions do not show any improved behaviour.
First of all \cite{RV} these actions are not 
on-shell tree-level improved in the sense of L\"uscher and Weisz \cite{LW}.
Indeed tree-level improvement requires $\alpha_2=0$ (see equations \reff{r1L}
and \reff{r2L}).
For $N=\infty$
it is also easy to see that these actions do not show any improved 
behaviour compared to the standard action \cite{CP_LAT96}.
Consider for instance the finite-size-scaling behaviour of the ratio
$\mu(\beta,\infty)/\mu(\beta,L)$. 
In table \reff{table_FSS} we report the value of 
the deviations from finite-size-scaling 
(see equation \reff{FSSformula})
\be
     R(L,x) = \left( {\mu(\beta,\infty)\over \mu(\beta,L)}\right)
      {1\over f_\mu(x)} -\, 1
\ee
for the various actions for $L\mu(\beta,L) \equiv x=2$ 
(this is the value of $x$ for which 
the deviations are larger). Clearly the action 
\reff{Symanzikdiag} has larger corrections that the standard action.
On the other hand,
as expected, the Symanzik actions \reff{SSymanzik} and \reff{Symanzik2}
show much smaller deviations from finite-size scaling, especially
 \reff{Symanzik2} which is Symanzik improved to all orders in perturbation 
theory and for which the corrections to finite-size scaling behave 
as $\log L/L^4$.
Notice that the improvement is working even for $L=4$--6 in spite 
of the large spatial extent of the Symanzik actions.
\begin{table}
\begin{center}
\begin{tabular}{|l|rrrr|}
\hline
$L$  & $S^{std}$   & $S^{diag}$  &  $S^{Sym}$  &  $S^{Sym2}$     \\
\hline
4  & 0.022022 &   0.039225  &  0.0002840   &  $-$0.0008378   \\
6  & 0.012721 &   0.021359  &  0.0007414   &  $-$0.0001953   \\
10 & 0.005758 &   0.009302  &  0.0003948   &  $-$0.0000246   \\
20 & 0.001806 &   0.002825  &  0.0001141   &  $-$0.0000011   \\
\hline
\end{tabular}
\end{center}
\caption{Values of 
$R(L;x)$ for $x=2$ for the various actions we have introduced in the 
text. Here $N=\infty$.
 }
\label{table_FSS}
\end{table}

We want now to show that if $\alpha_2\not=0$ an on-shell improved 
action can be obtained by adding a new four-spin coupling.

Indeed consider the continuum action 
\be
S = - N \int d^2 x \left[ {1\over2} (\bsigma\cdot \Box\bsigma) + 
  {\alpha_2 a^2\over2} (\bsigma\cdot \Box^2 \bsigma - 
    (\bsigma\cdot \Box \bsigma)^2 ) + O(a^4) \right]
\label{Honshellimp}
\ee
where $\Box = \sum_\mu \partial_\mu^2$. We will now show that, by means
of an isospectral transformation, one can get rid of the $O(a^2)$ 
terms\footnote{Equivalently one can show that the term proportional
to $a^2$ in \reff{Honshellimp} vanishes because of the equations
of motion \reff{eqmotion}.}.
Setting $\bsigma = (\bpi, \sqrt{1 - \bpi^2})$, the equation of motion
can be written as 
\be
\Box \bpi - \bpi (\bsigma\cdot \Box\bsigma) + O(a^2) = 0 \;\; .
\label{eqmotion}
\ee
Now consider the isospectral change of variable
\be
\bpi' = \bpi + 
     {A a^2\over2} (\Box \bpi - \bpi (\bsigma\cdot \Box\bsigma))\;\; .
\ee
where $A$ is a free parameter. It is immediate to verify that, setting 
$A= -\alpha_2$ we can rewrite 
\be
S = -{N\over2} \int d^2 x\, (\bsigma'\cdot \Box\bsigma') +\, 
O(a^4)
\ee
where $\bsigma' = (\bpi',\sqrt{1 - (\bpi')^2})$. As a final check 
we compute at one loop $\mu(\beta,L)$ for an action of the form
\reff{Honshellimp}. On the lattice we consider 
\be
S = N \sum_{xy} J(x-y) \bsigma_x \cdot \bsigma_y + 
    {\alpha_3 N\over2} \sum_{xyz}K(x-y) K(x-z) 
   (\bsigma_x \cdot \bsigma_y) (\bsigma_x\cdot \bsigma_z)
\label{Slattice-onshell}
\ee
where 
$\hat{K}(q) = q^2 + O(q^4)$ for $q\to 0$ and $\alpha_3$ is a free
parameter. For $a\to0$ the action \reff{Slattice-onshell} 
has the continuum limit \reff{Honshellimp} if 
$\alpha_1 = {1\over 12}$ and $\alpha_2 = \alpha_3$. 
Using the results of
\cite{RV,CPprep} we get the expansion \reff{LmuLpert} where
$r_1(L)$ and $r_2(L)$ are given by
\begin{eqnarray}
r_1(L) &=& - \int {d^2 k\over (2\pi)^2} {1\over w(k)} 
   \left[ {1\over4} \sum_\mu {\partial^2 w(k)\over \partial k_\mu^2} - 
   {\alpha_3\over2} 
    \sum_\mu \left({\partial \widehat{K} (k) \over \partial k_\mu}
      \right)^2 - 2 \alpha_3 \widehat{K}(k) - 
1\right] \nonumber \\ 
&& \qquad - {\pi\over 12 L^2} 
    (12 \alpha_1 + 8 \alpha_2 - 8 \alpha_3 -1) + O(L^{-4}) \;\; ,
\\
r_2(L) &=& {1\over 2\pi} \log L + \overline{F}_{00} + 
   \Lambda_0 + 2 \alpha_3 \int {d^2k\over (2 \pi)^2} 
   {\widehat{K}(k) \over w(k)} \nonumber \\
&& \qquad + {\pi\over 72 L^2} (1 - 12 \alpha_1) + O(L^{-4}) \;\; .
\end{eqnarray}
As expected the $1/L^2$ corrections vanish for $\alpha_1={1\over12}$ and 
$\alpha_2 = \alpha_3$.
An explicit example with all couplings lying in a plaquette is given by
\be
S^{onshell} = S^{diag} - {N\over24} \sum_x \sum_{i=1}^4 
     \left(\sum_{{\mu_i}} 
    (\bsigma_x\cdot\bsigma_{x+{\mu_i}} -1) \right)^2
\ee
where $\mu_1$ runs over the vectors $(1,0)$ and $(0,1)$,
$\mu_2$ over $(-1,0)$ and $(0,1)$,
$\mu_3$ over $(-1,0)$ and $(0,-1)$ and
$\mu_4$ over $(1,0)$ and $(0,-1)$ (the sum over $i$ symmetrizes 
over the four plaquettes stemming from the point $x$).
This action should be equivalent to $S^{Sym}$ for on-shell 
quantities. However $S^{onshell}$ enjoys an additional property: 
it is reflection positive for reflections with respect to the line 
$x_1 = 0$; it is then possible, with a standard construction, to 
define a transfer matrix 
\cite{reflectionpositivity1,reflectionpositivity2}.
It is also trivial to prove that this action has the correct continuum 
limit, in the sense that the ordered configuration is the unique 
absolute maximum of $S^{onshell}$.

The appearance of a four-spin coupling in $S^{onshell}$ suggests a connection 
between the on-shell improvement program and the perfect actions
\cite{perfect}. The perfect
action is indeed an action which is {\em tree-level} on-shell
improved\footnote{
It is somewhat misleading the statement \cite{Niedermayer_LAT96,Farchioni}
that perfect actions are {\em one-loop quantum perfect} because 
$\mu_2(L)$ has exponentially small corrections. 
Indeed cancellation of the $1/L^{2k}$ corrections in $\mu_2(L)$ fixes
the on-shell improved action only at tree-level \cite{LW,RV}.}
to all orders in $a^2$ \cite{Farchioni}. Work in this direction is in progress.

\bigskip 

We thank Ferenc Niedermayer, Giorgio Parisi, Paolo Rossi and Ettore Vicari for 
useful discussions.

\end{document}